/ 27 June 2000
\def\ptitle{\tiny Perturbation expansions for the spiked harmonic
oscillator
$\dots$}
\font\tr=cmr12                          
\font\bf=cmbx12                         
\font\sl=cmsl12                         
\font\it=cmti12                         
\font\trbig=cmbx12 scaled 1500          
\font\tiny=cmr10                        

\output={\shipout\vbox{\makeheadline
                                      \ifnum\the\pageno>1 {\hrule}  \fi
                                      {\pagebody}
                                      \makefootline}
                   \advancepageno}

\headline{\noindent {\ifnum\the\pageno>1
                                   {\tiny \ptitle\hfil
page~\the\pageno}\fi}}
\footline{}

\tr
\def\bra{{\rm <}}    
\def\ket{{\rm >}}    
\baselineskip 15 true pt  
\parskip=0pt plus 5pt
\parindent 0.25in
\hsize 6.0 true in
\hoffset 0.25 true in
\emergencystretch=0.6 in           
\vfuzz 0.4 in                            
\hfuzz  0.4 in                           
\vglue 0.1true in
\mathsurround=2pt                        
\topskip=24pt                            
\newcount\zz  \zz=0  
\newcount\q   
\newcount\qq    \qq=0  
\def\pref #1#2#3#4#5{\frenchspacing \global \advance \q by 1         
\edef#1{\the\q}
       {\ifnum \zz=1 { %
         \item{[\the\q]}
         {#2} {\bf #3},{ #4.}{~#5}\medskip} \fi}}
\def\bref #1#2#3#4#5{\frenchspacing \global \advance \q by 1
    \edef#1{\the\q}
    {\ifnum \zz=1 { %
       \item{[\the\q]}
       {#2}, {\it #3} {(#4).}{~#5}\medskip} \fi}}

\def\gref #1#2{\frenchspacing \global \advance \q by 1  
    \edef#1{\the\q}
    {\ifnum \zz=1 { %
       \item{[\the\q]}
       {#2}\medskip} \fi}}

\def\sref #1{~[#1]}

\def\references#1{\zz=#1
   \parskip=2pt plus 1pt   
   {\ifnum \zz=1 {\noindent \bf References \medskip} \fi} \q=\qq

\pref{\agub}{V. C. Aguilera-Navarro and R. Guardiola, J. Math.
Phys.}{32}{2135 (1991)}{}

\pref{\esta}{E. S. Est\'evez-Bret\'on and G. A.
Est\'evez-Bret\'on,  J. Math. Phys.}{34}{437 (1993)}{}

\pref{\znob}{M. Znojil, J. Math. Phys.}{34}{4914 (1993)}{}

\pref{\hald}{R. Hall, N. Saad and A. von Keviczky, J. Math.
Phys.}{39}{6345-51 (1998)}{}

\bref{\gold}{I. I. Gol'dman and D. V. Krivchenkov}{Problems
in Quantum mechanics}{Pergamon, London, 1961}{}

\bref{\hys}{J. M. Hyslop}{Infinite Series}{Interscience
Publishers, New York, 1959}{}

\bref{\gra}{I.S. Gradshteyn and I.M. Ryzhik}{Table of Integrals, Series,
and Product, $5^{th}$ Edition}{Academic Press, New York, 1994}{}

\pref{\halda}{R. Hall and N. Saad, J. Phys. A}{}{In Press}{}

\bref{\slat}{L. J. Slater}{Confluent Hypergeometric
Functions}{University Press, Cambridge, 1960}{}

\bref{\abr}{M. Abramowitz and I. A. Stegun}{Handbook of
Mathematical Functions}{Dover, New York, 1970}{}

\pref{\mit}{K. Mitchell, Phil. Mag.}{40}{351 (1949)}{}

\bref{\ageb}{V. C. Aguilera-Navarro and E. S. Est\'evez-Bret\'on}
{Sums of infinite series involving gamma functions}
{Private communication}{}
} 

\references{0}    
\centerline{\bf\trbig Perturbation expansions for the spiked harmonic
oscillator}
\centerline{\bf\trbig and related series involving the gamma function}
\medskip
\vskip 0.5 true in
\centerline{Richard L. Hall and Nasser Saad}
\medskip
{\leftskip=0pt plus 1fil
\rightskip=0pt plus 1fil\parfillskip=0pt\baselineskip 18 true pt
\obeylines
Department of Mathematics and Statistics, Concordia University,
1455 de Maisonneuve Boulevard West, Montr\'eal,
Qu\'ebec, Canada H3G 1M8.\par}
\vskip 0.25 true in

\vskip 0.5 true in

\bigskip
\centerline{\bf Abstract}\medskip

We study weak-coupling perturbation expansions for the ground-state energy of the Hamiltonian with the 
generalized
spiked harmonic oscillator potential $V(x)=Bx^2+{A\over
x^2}+{\lambda\over x^\alpha}$, and also for the bottoms of the angular
momentum subspaces labelled by $l=0,1,\dots,$ in $N$-dimensions corresponding to 
the spiked harmonic oscillator potential
$V(x)=x^2+{\lambda\over
x^\alpha},$ where $\alpha$ is a real positive parameter. A method of
Znojil is then applied to obtain closed form
expressions for the sums of some infinite series whose terms involve
ratios and products of gamma functions.

\vfill\eject

\noindent{\bf I. Introduction}\medskip
The spiked harmonic oscillator Hamiltonian
defined by
$$H=-{d^2\over dx^2}+x^2+{\lambda\over x^\alpha},\ \alpha<5/2,\ x\in
[0,\infty)\eqno(1.1)$$
where the positive parameter $\lambda$ measures the strength
of the singular term, has been the subject of intensive study\sref{\agub-\znob}. Aguilera-Navarro and Guardiola\sref{\agub},
employed a resummation technique to obtain a weak coupling perturbation
expansions for the ground state energy of the Hamiltonian
(1.1), using standard perturbation theory up to second order. The Hamiltonian 
(1.1) is first written $H=H_0+\lambda V;$ then,
using the odd-parity solutions of the one-dimensional harmonic oscillator 
$\psi_n(x)=|n\ket$ satisfying the Dirichlet boundary
condition
$\psi(0)=0,$ and the unperturbed energies  $E_n=3+4n,\ n=0,1,2,\dots$,
they found that the weak-coupling expansion for the ground state of $H$ to the
second order in $V$ becomes
$$
E=E_0+\lambda\bra 0|x^{-\alpha}|0\ket+\lambda^2\sum\limits_{n\geq 1}
{{|\bra 0|x^{-\alpha}|n\ket|^2}\over E_0-E_n}+\dots,\ \alpha <
5/2,\eqno(1.2)
$$
where $\bra 0|x^{-\alpha}|n\ket$ given by
$$
\bra 0|x^{-\alpha}|n\ket=
{{(-2)^n}\over {\sqrt{(2n+1)!}}}{{\Gamma({3-\alpha\over
2})\Gamma({\alpha\over 2}+n)}\over {\Gamma({3\over
2})\Gamma({\alpha\over
2})}}\quad n=0,1,\dots.\eqno(1.3)
$$
In their pertubation treatment, Aguilera-Navarro and Guardiola
construct a function $F$, expressed in terms of a particular form of
the Gauss hypergeometric series; the function $F$ was found to be useful
for obtaining analytic approximations for the ground state energy of the for the 
`non-supersingular' cases: $\alpha={1\over
2}$,$\alpha=1$, and
$\alpha={3\over 2}$. They also provided a weak coupling expression valid
for the case $\alpha=2. $ Est\'evez-Bret\`on et al\sref{\esta}
derived an exact analytical result, by using the
function $F,$ valid for the special case $\alpha=2$.  Later Znojil\sref{\znob}
derived the same result by  an elegant and economical method. It is 
perhaps worth noting here that in all these
works, although the conclusions and the results were correct, there remained an 
error in the $F$
formula, Eq.(14) in Ref.[1], deduced by Aguilera-Navarro and Guardiola,
which should read
$$F={1\over 8({\alpha\over 2}-1)^2}\bigg[{}_2F_1\bigg({\alpha\over
2}-1,{\alpha\over 2}-1;{1\over 2};1\bigg)-1-
{2({\alpha\over 2}-1)^2}\bigg].$$
In Sec. II of the present paper we generalize the weak coupling
expansion (1.2)
to study the generalized spiked harmonic oscillator Hamiltonian
$$H\equiv H_0+\lambda V=-{d^2\over dx^2}+Bx^2+{A\over x^2}+{\lambda\over
x^\alpha},\quad \alpha<5/2,\ x\in [0,\infty).
\eqno(1.4)$$
where $\lambda$ and $\alpha$ are positive parameters. We show that
the weak coupling expansion, in this case, is given by
$$
E=2\sqrt{B}\gamma+B^{\alpha\over 4}{{\Gamma(\gamma-{\alpha\over
2})}\over \Gamma(\gamma)}\lambda-\lambda^2
{{B^{\alpha-1\over 2}\alpha^2}\over 16\gamma}
{{\Gamma^2(\gamma-{\alpha\over 2})}\over
\Gamma^2(\gamma)}\ {}_4F_{3}(1,1,{\alpha\over 2}+1,{\alpha\over
2}+1;\gamma+1,2,2;1)+\dots,\quad $$
where $\gamma=1+{1\over 2}\sqrt{1+4A}$. This expression is valid for all 
values of $\alpha<\gamma+1,$ including
$\alpha=2$. This formula allows us to obtain perturbation expansions for (1.1) and (1.4)
valid for the bottoms of the angular momentum subspaces labeled by
$l=0,1,\dots$ in $N$-dimensions. In Sec.III, we adopt the constructive
approach of Aguilera-Navarro and Guardiola to generalized the function
$F$ which we then use to obtain a sum for the infinite series
$$\sum\limits_{n\geq 1}{({\alpha\over 2})_n^2\over 4n(n+1)\ (\gamma)_n\
n!}$$
In Sec.IV, by employing Znojil's technique\sref{\znob}, our generalization
turns out to be useful for obtaining closed-form sums for other
interesting infinite series whose terms involve ratios and products of gamma
functions.
\medskip
\noindent{\bf II. Weak coupling expansions}
\medskip
Recently, we have obtained expressions\sref{\hald} for the
singular-potential integrals $\bra m|x^{-\alpha}|n\ket$ of the
Hamiltonian (1.4) using the
Gol'dman and Krivchenkov eigenfunctions\sref{\gold}
$$
\left\{
\eqalign{
&\psi_n(x)\equiv |n\ket =C_nx^{{1\over 2}(1+\sqrt{1+4A})}e^{-{1\over
2}\sqrt{B} x^2}{}_1F_1(-n,1+{1\over 2}\sqrt{1+4A};\sqrt{B}x^2);\cr
&C_n^2=
{{2B^{{1\over 2}+{1\over 4}\sqrt{1+4A}}\Gamma(n+1+{1\over
2}\sqrt{1+4A})}
\over
{{n![\Gamma(1+{1\over 2}\sqrt{1+4A}})]^2}},\quad
n=0,1,2,\dots,\cr}\right.\eqno(2.1)
$$
where ${}_1F_1$ is the confluent hypergeometric function\sref{\slat}
$$
{}_1F_1(a,b;z)=\sum\limits_k {{(a)_kz^k}\over {(b)_kk!}},\
(a)_k=a(a+1)\dots(a+k-1)={{\Gamma(a+k)}\over {\Gamma(a)}}.\eqno(2.2)
$$
and the exact eigenenergies $E_n=\sqrt{B}(4n+2+\sqrt{1+4A}),\
n=0,1,2,\dots,$ for the singular Hamiltonian
$$H_0=-{d^2\over dx^2}+Bx^2+{A\over x^2},\quad B>0,A\ge 0,\ x\in
[0,\infty).
\eqno(2.3)$$
Hall et al found\sref{\hald}, for $\alpha<2\gamma$, that the matrix
elements $\bra m|x^{-\alpha}|n\ket$ are given by
$$
\eqalign{\bra m|x^{-\alpha}|n\ket&=(-1)^{n+m}B^{\alpha/4}
\sqrt{{\Gamma{(\gamma+m)}}\over {n!m!\ \Gamma{(\gamma+n)}}}\cr
&\times \sum_{k=0}^m (-1)^k{m \choose k}{{\Gamma(k+\gamma-{\alpha\over
2})\Gamma({\alpha\over 2}-k+n)}\over
{\Gamma(k+\gamma)\Gamma({\alpha\over
2}-k)}},\ \gamma=1+{1\over 2}\sqrt{1+4A},\ \cr}\eqno(2.4)
$$
in which each element has a factor which is a polynomial of degree $m+n$ in 
$\alpha$. The relevant matrix elements $\bra
0|x^{-\alpha}|n\ket$ are given\sref{\hald} by
$$\bra 0|x^{-\alpha}|n \ket=(-1)^{n}B^{\alpha/4}
\sqrt{{\Gamma{(\gamma)}}\over {n!\
\Gamma{(\gamma+n)}}}{{\Gamma(\gamma-{\alpha\over 2})\Gamma({\alpha\over
2}+n)}\over {\Gamma(\gamma)\Gamma({\alpha\over 2})}},\
n=0,1,2,\dots,\eqno(2.5)$$
Thus, writing (1.4) as $H=H_0+\lambda V$ and using the standard
perturbation theory to the second order, we find that the weak coupling
expansion (1.2) now reads
$$E=2\sqrt{B}\gamma+B^{\alpha\over 4}{{\Gamma(\gamma-{\alpha\over
2})}\over \Gamma(\gamma)}\lambda-\lambda^2B^{\alpha-1\over
2}{{\Gamma^2(\gamma-{\alpha\over 2})}\over
\Gamma^2(\gamma)}\sum\limits_{n\geq 1}{({\alpha\over 2})_n^2\over 4n\
(\gamma)_n\ n!}+\dots,\ \alpha<\gamma+1\eqno(2.6)$$
where $\gamma=1+{1\over 2}\sqrt{1+4A}$. We observe that the ratio of the
$n$th and $(n+1)$th terms of the sum in the coefficient of $\lambda^{2}$ in (2.6) is
$${{\bra 0|x^{-\alpha}|n\ket^2/(E_n-E_0)}\over {\bra
0|x^{-\alpha}|n+1\ket^2/(E_{n+1}-E_0)}}=1+{{\gamma+2-\alpha}\over
n}+o({1\over n^2}),\quad \sl{as}\quad n\rightarrow \infty,$$
so that, by Raabe's test\sref{\hys}, this sum is convergent for
$\alpha<\gamma+1$. The expressions (1.2) and (2.6) are accurate for
$\lambda$ small compared to unity. It is interesting that the sum
of the infinite series in the $\lambda^2$ coefficient can be
computed exactly for arbitrary values of $\alpha$ and $\gamma$
satisfying $\alpha<\gamma+1$. We express the sum in terms of the
generalized hypergeometric functions ${}_pF_q$ defined\sref{\gra} by
$${}_pF_q(\alpha_1,\alpha_2,\dots,\alpha_p;\beta_1,\beta_2,\dots,\beta_q;z)=\sum
\limits_{n=0}^{\infty}{{(\alpha_1)_n(\alpha_2)_n\dots(\alpha_p)_n}\over{(\beta_1
)_n(\beta_2)\dots(\beta_q)_n}}{z^n\over
n!}.\eqno(2.7)$$
Indeed, the sum in the $\lambda^2$ coefficient implies
$$\eqalign{\sum\limits_{n\geq 1}{({\alpha\over 2})_n^2\over 4n\
(\gamma)_n\ n!}&=\sum\limits_{n\geq 1}{{(n-1)!({\alpha\over
2})_n^2}\over {4\
(\gamma)_n\ (n!)^2}}\cr
&={1\over 4}\sum\limits_{n=0}{{(n!)^2({\alpha\over 2})_{n+1}^2}\over {
\ ((n+1)!)^2\ (\gamma)_{n+1}\ n!}}\cr
&={\alpha^2\over 16\gamma}\sum\limits_{n=0}{{(1)_n^2({\alpha\over
2}+1)_{n}^2}\over {
\ (2)_n^2\ (\gamma+1)_{n}\ n!}}\cr
&={\alpha^2\over 16\gamma}\ {}_4F_{3}(1,1,{\alpha\over 2}+1,{\alpha\over
2}+1;\gamma+1,2,2;1)\cr}\eqno(2.8)
$$
where we have used the Pochhammer identities $(z)_{n+1}=z(z+1)_n$ and
$n!=(1)_n$.
This expression can easily computed for arbitrary values of
$\alpha<\gamma+1$
by the use, for example, of Mathematica.  The weak coupling expansion (2.6) now
reads
$$
E=2\sqrt{B}\gamma+B^{\alpha\over 4}{{\Gamma(\gamma-{\alpha\over
2})}\over \Gamma(\gamma)}\lambda-\lambda^2
{{B^{\alpha-1\over 2}\alpha^2}\over 16\gamma}
{{\Gamma^2(\gamma-{\alpha\over 2})}\over
\Gamma^2(\gamma)}\ {}_4F_{3}(1,1,{\alpha\over 2}+1,{\alpha\over
2}+1;\gamma+1,2,2;1)+\dots.\eqno(2.9)$$
The results of Aguilera-Navarro and Guardiola for the special case
$B=1$, $A=0$ or $\gamma=3/2$, and for the values of $\alpha={1\over
2}$,$\alpha=1$,$\alpha={3\over 2}$, and $\alpha=2$ follow immediately
without the necessity of special treatment for the case of $\alpha=2$ as
suggested before by many workers in the field\sref{\agub-\znob}.
The expression (2.9) can be further generalized  to apply to the ground
state eigenenergy at the bottom of each angular momentum subspace
labeled by $l=0,1,2,\dots$ in $N$-dimensions: we just need\sref{\halda}
 to replace $A$ with $A\rightarrow A+(l+1/2(N-1))(l+1/2(N-3)).$
For the spiked harmonic oscillator potential (1.1), we set $A=0$ or we
replace $\gamma$ with $l+{N\over 2}$ to obtain a weak-coupling
expansion valid for the bottoms of the angular-momentum subspaces in
$N$-dimensions.
\medskip
\noindent{\bf III. The $F$ function}
\medskip
Although, our results in section (II) cover all the cases for
$\alpha<5/2$ for the Hamiltonians (1.1) and (1.4) the
constructive approach of Aguilera-Navarro and Guardiola allows us to
obtain more sums of infinite series involving gamma
functions.  We first  generalized the function $F$ as
introduced in Ref.\sref{\agub} (we also point out the error in
the $F$ formula there). If we denote the sum in the $\lambda^{2}$
coefficient, of Eq.(2.6), by
$$
G=\sum\limits_{n\geq 1}{({\alpha\over 2})_n^2\over 4n\ (\gamma)_n\
n!},\eqno(3.1)
$$
and compare this with the sum
$$
F=\sum\limits_{n\geq 1}{({\alpha\over 2})_n^2\over {4(n+1)\ (\gamma)_n\
n!}}.\eqno(3.2)
$$
We see that $G$ and $F$ are related by the expression
$$
G=F+\sum\limits_{n\geq 1}{({\alpha\over 2})_n^2\over 4n(n+1)\
(\gamma)_n\
n!}.\eqno(3.3)
$$
The new expression for the sum thus obtained will be easier to
approximate since fewer terms will be required for a given accuracy. Moreover, 
using the Pochhammer identity
$(z)_{n+1}=z(z+1)_n$, we note that $F$ can be
written in terms of a special form of the Gauss hypergeometric function\sref{\slat}
$${}_2F_{1}(a,b;c;z)=\sum\limits_{n=0}{{(a)_n(b)_n}\over (c)_n}{z^n\over
n!}\eqno(3.4)$$
(with a circle of convergence $|z|=1$) as
$$F={(\gamma-1)\over 4({\alpha\over
2}-1)^2}\bigg[{}_2F_1\bigg({\alpha\over
2}-1,{\alpha\over 2}-1;{\gamma-1};1\bigg)-1-
{ {({\alpha\over 2}-1)^2}\over (\gamma-1)}\bigg]
.\eqno(3.5)$$
This generalizes the weak coupling expansion derived by
Aguilera-Navarro and Guardiola to study Eq.(1.4) for $\alpha<\gamma+1$.
However, we should note here the correct form of function $F$ of the
case $\gamma=3/2$ or $A=0$ that is
$$F={1\over 8({\alpha\over 2}-1)^2}\bigg[{}_2F_1\bigg({\alpha\over
2}-1,{\alpha\over 2}-1;{1\over 2};1\bigg)-1-
{2({\alpha\over 2}-1)^2}\bigg],\eqno(3.6)$$
not as quoted in\sref{\agub-\znob}.
Eq.(2.8) and (3.5) can be used now to obtain a sum for the
infinite series
$$\eqalign{\sum\limits_{n\geq 1}{({\alpha\over 2})_n^2\over 4n(n+1)\
(\gamma)_n\
n!}={\alpha^2\over 16\gamma}&\ {}_4F_{3}(1,1,{\alpha\over
2}+1,{\alpha\over 2}+1;\gamma+1,2,2;1)\cr
&-{(\gamma-1)\over 4({\alpha\over
2}-1)^2}\bigg[{}_2F_1\bigg({\alpha\over
2}-1,{\alpha\over 2}-1;{\gamma-1};1\bigg)-1-
{ {({\alpha\over 2}-1)^2}\over (\gamma-1)}\bigg]\cr}\eqno(3.7)$$
valid for $\alpha\neq 2$.
For the special limit $\alpha=2$, the expression (3.5) has no meaning.
However, the sum in the $\lambda^2$ coefficient of (2.6) converges and
follows from (2.8) by setting $\alpha=2$. Indeed, in this case, the
sum in (2.6) becomes
$$\sum\limits_{n\geq 1}{(1)_n^2\over
{4n\ (\gamma)_n\ n!}}=\sum\limits_{n\geq 1}{(n-1)!\over
{4\ (\gamma)_n }}=\sum\limits_{n=0}{(1)_n^2\over
{4\ (\gamma)_{n+1}\ n!}}={1\over 4\gamma}\
{}_2F_{1}(1,1;\gamma+1;1),\eqno(3.8)
$$
which follows immediately from (2.8). The Gauss hypergeometric function
${}_2F_{1}(1,1;\gamma+1;1)$ can be evaluated using the identity\sref{\slat}
$${}_2F_1(a,b,c;1)={{\Gamma(c)\Gamma(c-a-b)}\over
{\Gamma(c-a)\Gamma(c-b)}}, \quad c-a-b>0,c>b>0\eqno(3.9)$$
to obtain
$$\sum\limits_{n\geq 1}{(1)_n^2\over
{4n\ (\gamma)_n\ n!}}={1\over 4(\gamma-1)},\quad \gamma>1.\eqno(3.10)$$
Thus, for the case $\alpha=2$, the weak coupling expansion (2.9) becomes
$$E(\alpha=2)=2\sqrt{B}\gamma+{\sqrt{B}\over
(\gamma-1)}\lambda-{\sqrt{B}\over 4(\gamma-1)^3}\lambda^2
+\dots,\ \gamma>1\eqno(3.11)
$$
and finally, for $B=1$, $A=0$ or $\gamma=3/2$, we get
$$E(\alpha=2)=3+2\lambda-2\lambda^2+\dots,\ \eqno(3.12)
$$
as we expect.
\medskip
\noindent{\bf IV. More closed-form sums of infinite series}
\medskip
For the special limiting case $\alpha\rightarrow 2$, we introduce a
parameter
$\epsilon={\alpha\over 2}-1$ which will be chosen to approach zero.  The
function $F$,
as given by (3.5), becomes in this limit
$$\lim\limits_{\epsilon\rightarrow 0} F={{(\gamma-1)}\over
4}\lim\limits_{\epsilon\rightarrow
0}\epsilon^{-2}\bigg[{}_2F_{1}(\epsilon,\epsilon,\gamma-1,1)-1\bigg]-{1\over
4}.\eqno(4.1)
$$
Using the series expansion of the Gauss hypergeometric function (3.4),
Eq.(4.1) can be written, using $\Gamma(z+1)=z\Gamma(z)$, as
$$\lim\limits_{\epsilon\rightarrow 0} F={1\over 4}\sum\limits_{n=
0}^\infty{{\Gamma(n)\Gamma(\gamma)}\over n\Gamma(n+\gamma-1)}-{1\over
4}.\eqno(4.2)$$
\noindent Some results similar to Eq.(4.1) and (4.2), were first published, 
without detailed proofs, by Mitchell\sref{\mit}.
Now, using the identity (3.9),
Eq. (4.1) can now be rewritten as
$$\lim\limits_{\epsilon\rightarrow 0} F={{(\gamma-1)}\over
4}\lim\limits_{\epsilon\rightarrow
0}\epsilon^{-2}\bigg[{{\Gamma(\gamma-1)\Gamma(\gamma-1-2\epsilon)}\over
{\Gamma^2(\gamma-1-\epsilon)}}-1\bigg]-{1\over 4}.\eqno(4.3)
$$
Now employing Znojil's method \sref{\znob}, we can obtain closed form sums for 
other infinite series involving Gamma
functions.
Indeed, the Maclaurin's expansion of the gamma function
$$
\Gamma(c+x)=
\Gamma(c)\bigg\{1+x\psi(c)+{1\over
2}[\psi^2+\psi^{(1)}(c)]+\dots\bigg\}\eqno(4.4)
$$
where $\psi(c)$ and $\psi^{(n)}(c)$,${n\ge 1}$, are the digamma and the
polygamma functions\sref{\abr}, respectively. We can show, after
expanding the gamma functions in (4.3) and employing multiplication and
division of polynomials, that Eq.(4.4) can be written as
$$
\lim\limits_{\epsilon\rightarrow 0} F= {(\gamma-1)\over
4}\psi^{(1)}(\gamma-1)-{1\over 4}\eqno(4.5)$$
where $\psi^{(n)}(z)$ are the polygamma functions.
Comparing (4.2) and (4.5), we have
$$\sum\limits_{n=1}^\infty{{\Gamma(n)\Gamma(\gamma-1)}\over
n\Gamma(n+\gamma-1)}=\psi^{(1)}(\gamma-1), \quad \gamma>1,\eqno(4.6)$$
where $\psi^{(1)}(z)$ is the trigamma function\sref{\gra}.
For the purpose principally of verification we now note some special cases. For 
$\gamma=2$
$$
\sum\limits_{n=1}^\infty{1\over {n^2}}={\pi^2\over 6},\eqno(4.7)
$$
and for $\gamma-1=m\ge 2$ positive integer, we find
$$
\sum\limits_{n=1}^\infty{{\Gamma(n)\Gamma(m)}\over
{n\Gamma(n+m)}}={\pi^2\over 6}-\sum\limits_{k=2}^m{1\over
(k-1)^2},\eqno(4.8)$$
by using the recurrence relation
$$\psi^{(n)}(z+1)=\psi^{(n)}(z)+(-1)^{n}n!z^{-n-1}.$$
Further, for $\gamma=3/2$
$$
\sum\limits_{n=1}^\infty{{\Gamma(n)\Gamma(1/2)}\over
{n\Gamma(n+1/2)}}={\pi^2\over 2}.\eqno(4.9)$$
and
$$
\sum\limits_{n=1}^\infty{{\Gamma(n)\Gamma(mz)}\over
{n\Gamma(n+mz)}}={1\over m^2}\sum\limits_{k=0}^{m-1}\psi^{(1)}(z+{k\over
m}).\eqno(4.10)$$
Finally, we can now have a finite sum for the infinite series
(3.7) for the case $\alpha=2$ and $\gamma>1$,
$$
\sum\limits_{n\geq 1}{({1})_n^2\over 4n(n+1)\ (\gamma)_n\
n!}=
{1\over 4\gamma}\ {}_2F_{1}(1,1;\gamma+1;1)-{{\gamma-1}\over
4}\psi^{(1)}(\gamma-1)+{1\over 4}.\eqno(4.11)$$
\medskip
\noindent{\bf V. Conclusion}
\medskip
We have obtained a compact weak-coupling expansion (2.9) for eigenvalues of the 
spiked harmonic oscillator Hamiltonian.  Our
expansion extends the earlier work of Aguilera-Navarro and Guardiola for 
$\gamma\neq 3/2,$ and it allows for arbitrary spatial
dimension $N$ and also, for $N \geq 2,$ arbitrary  orbital angular momentum 
$\ell.$ Moreover, with the closed-form expressions
we have been able to provide for the coefficient of the $\lambda^2$ term, the 
new expansion is easier to handle and calculate
with, even at or near to the special value $\alpha=2$. These analytic expressions describe approximately how the eigenvalues depend on all the parameters in the Hamiltonian.  Such formulas are complementary to data obtained with the aid of a computer; moreover, they are useful in guiding a procedure that searches for very accurate numerical eigenvalues.  As a byproduct of this work, we have been led to some simple closed-forms for a variety of interesting
infinite series involving sums and ratios of gamma functions.\medskip
\noindent{\bf Acknowledgments}\medskip
Partial financial support of this work under Grant No. GP3438 from the Natural 
Sciences and Engineering Research Council of
Canada is gratefully acknowledged by one of us (RLH).  We should also like to 
thank Professors V. C. Aguilera-Navarro and G.A.
Est\'evez-Bret\'on for discussion through a private communication\sref{\ageb}.

\vfil\eject
\references{1}
\end